\documentstyle[12pt]{article}

\newlength{\titlesep}
\newlength{\authorsep}
\makeatletter

\renewcommand{\thesection}{\Roman{section}}

\def\fnum@figure{FIG.~\thefigure}

\newcommand{\reffig}[1]{Fig.~\protect\ref{#1}}

\newcounter{figureparent}
\@addtoreset{figure}{figureparent}

\newcounter{eqnparent}
\@addtoreset{equation}{eqnparent}

\renewcommand{\abstract}{\if@twocolumn
  \section*{Abstract}
  \else
  \begin{center}
    {\bf Abstract\vspace{-.5em}\vspace{0pt}}
  \end{center}
  \fi}
\renewcommand{\endabstract}{\if@twocolumn\else\endquotation\fi}

\renewcommand{\appendix}{\par
  \setcounter{section}{0}
  \setcounter{subsection}{0}
  \renewcommand{\thesection}{Appendix~\Alph{section}}
  \renewcommand{\theequation}{(\Alph{section}.\arabic{equation})}}

\newcommand{\thismonth}{\ifcase\month\or
 January\or February\or March\or April\or May\or June\or
 July\or August\or September\or October\or November\or December\fi
 \space \number\year}

\newcommand{\preprintnumber}[1]
{\begin{flushright}
  \begin{tabular}{l} #1 \end{tabular}
  \end{flushright}}

\makeatother

\newcommand{\Rn}[1]{{\uppercase\expandafter{\romannumeral#1}}}
\newcommand{\gsim}%
{\mathrel{\mbox{\raisebox{-1.0ex}%
{$\stackrel{\textstyle >}{\textstyle \sim}$}}}}
\newcommand{\lsim}%
{\mathrel{\mbox{\raisebox{-1.0ex}%
{$\stackrel{\textstyle <}{\textstyle \sim}$}}}}
\newcommand{\ie}{{\it i.e.\/}\ }
\newcommand{\br}{{\rm Br}}

\newcommand{\Journal}[4]{{#1} {\bf #2} {(#3)} {#4}}
\newcommand{\pl}{\sl Phys.~Lett.}
\newcommand{\plb}{\sl Phys.~Lett.~{\bf B}}

\newcommand{\pr}{\sl Phys.~Rev.}
\newcommand{\prd}{\sl Phys.~Rev.~{\bf D}}
\newcommand{\prl}{\sl Phys.~Rev.~Lett.}
\newcommand{\np}{\sl Nucl.~Phys.}
\newcommand{\npb}{\sl Nucl.~Phys.~{\bf B}}
\newcommand{\ptp}{\sl Prog.~Theor.~Phys.}

\newcommand{\zpc}{\sl Z.~Phys.~{\bf C}}

\newcommand{\epsfile}[1]{\relax}

\begin{document}
\baselineskip 18pt

\begin{titlepage}
\preprintnumber{%
KEK-TH-484 \\
KEK Preprint 97-3\\
OU-HET 261 \\
TU-519 \\
April 1997
}
\vspace*{\titlesep}
\begin{center}
{\LARGE\bf
$b\rightarrow s\,\ell\overline{\ell}$ process and multi-Higgs doublet model
}
\\
\vspace*{\titlesep}
{\large Yasuhiro  Okada$^1$}\\
{\large Yasuhiro  Shimizu$^2$},\\
and
{\large Minoru  Tanaka$^3$}\\
\vspace*{\authorsep}
{\it $^1~ ^,$ $^2$Theory Group, KEK, Tsukuba, Ibaraki, 305 Japan }
\\
\vspace*{\authorsep}
{\it $^2$Department of Physics, Tohoku University \\
  Sendai 980-77 Japan}
\\
\vspace*{\authorsep}
{\it $^3$ Department of Physics, Osaka University \\
  Toyonaka, Osaka, 560 Japan }
\end{center}
\vspace*{\titlesep}
\begin{abstract}
Rare $b$ decay processes are analyzed in the multi-Higgs doublet model.
Taking account of the constraint from the $b\rightarrow s\,\gamma$ 
process, the branching ratio and the forward-backward
asymmetry of the final leptons  for 
the $b\rightarrow s\,\ell^+\ell^-$ process are calculated.
It is shown that the branching ratio can be a few times 
larger than the standard model prediction and the asymmetry
can be significantly different from that in the standard model.
Combining these observable quantities it is possible to determine 
complex coupling constants associated with
the charged Higgs mixing matrix. 
The CP violating charge 
asymmetry in the $b\rightarrow s\,\ell^+\ell^-$ process 
and the branching ratio of the $b\rightarrow s\,\nu\overline{\nu}$
process are also calculated.
\end{abstract}
\end{titlepage}
In search for new physics beyond the Standard Model (SM),
flavor changing neutral current (FCNC) processes can play an
important role. Since the source of the flavor changing processes only
lies in the Cabbibo-Kobayashi-Maskawa (CKM) matrix in the SM, it is possible
to make precise predictions for various observable quantities 
in FCNC processes within the SM.
An important example is the branching ratio
of the $b\rightarrow s\,\gamma$ process, which was reported to be
$\br(b\rightarrow s\,\gamma) = (2.32\pm0.57\pm0.35)\times 10^{-4}$ from
the CLEO experiment \cite{CLEO}. Using the observed top quark mass, 
the measured branching ratio is consistent with the SM prediction. 
Therefore this process becomes a very strong constraint on models beyond
the SM such as two Higgs doublet model \cite{2HDM} and supersymmetric 
extensions of the SM \cite{MSSM}. 
In addition to the $b \rightarrow s\,\gamma$ process, the
$b\rightarrow s\,\ell^+ \ell^-$ and $b \rightarrow s\,\nu\overline{\nu}$
processes can be important 
constraints on new physics. Current experimental bounds for these processes
are only by one order of magnitudes above the SM predictions
\cite{bsll,bsnn}.

In this letter we consider possible constraints on parameters 
in the multi-Higgs doublet model from these rare $b$ decay processes.
When the number of the Higgs doublets is more than three we can
introduce complex phases in the charged Higgs mixing matrix.
In this sense the multi-Higgs doublet model is a natural extension of 
the SM which involves
a new source of the CP violation. Present phenomenological constraints
on coupling constants in this model are considered, for example, in 
Ref.\cite{MHDM}. 
Remarkably it is pointed out in Ref.\cite{GN} that the strongest
constraint on the imaginary part of coupling constant comes from the 
$b \rightarrow s\,\gamma$ process, not from the CP violating quantities
such as neutron electric dipole moment.
Here we study the 
$b \rightarrow s\,\ell^+\ell^-$ and 
$b \rightarrow s\,\nu \bar{\nu}$ processes in the same model. 
We found that within the present experimental constraints including the
$b \rightarrow s\,\gamma$ process the branching ratio
can be a few times larger than the
SM predictions and the forward-backward asymmetry of the final leptons 
for $b \rightarrow s\,\ell^+\ell^-$ process can significantly differ
from the SM.
Combining information from the branching ratio and the forward-backward
asymmetry, it is possible to determine the complex coupling constants in 
future experiments.
We also calculate the CP violating charge asymmetry
between the $b \rightarrow s\,\ell^+\ell^-$ and 
${\overline b} \rightarrow {\overline s}\,\ell^+\ell^-$ decays,
which turns out to be a few percent in the allowed region.

We consider the multi-Higgs doublet model with
the following Yukawa couplings,

\begin{equation}
  \label{lagrangian}
  {\cal L} = {\overline q}_L\,y_d\,d_R\,H_d 
           + {\overline q}_L\,y_u\,u_R\,H_u
           + {\overline \ell}_L\,y_{\ell}\,e_R\,H_{\ell}
           + h.c. ,
\end{equation}
where $y_d, y_u, y_{\ell}$ are $3 \times 3$ matrices. We have assumed
that up-type quarks, down-type quarks and leptons have Yukawa couplings
with only one Higgs doublet, $H_u, H_d, H_{\ell}$ respectively. 
In this way we can avoid FCNC effects at the tree level \cite{NFC}.
The couplings of physical charged Higgs bosons with fermions are then 
given by
\begin{eqnarray}
  \label{chargedhiggs}
  {\cal L} = (2\sqrt{2}G_F)^{1/2} \sum_{i=1}^{n-1}
             ( X_i\,{\overline u}_L\,V\,M_D\,d_R\,H_i^+
             + Y_i\,{\overline u}_R\,M_U\,V\,d_L\,H_i^+
             + Z_i\,{\overline \nu}_L\,M_E\,e_R\,H_i^+ ) ,
\end{eqnarray}
where $n$ is the number of Higgs doublets, $H_i^+ (i=1\sim n-1)$
represent mass eigen states of charged Higgs bosons and
$V$ is the CKM matrix.
New CP violating complex phases arise in the charged Higgs 
mixing matrix if there are three or more Higgs doublets.
In such cases the coupling constants $X_i$, $Y_i$, $Z_i$ are in general 
complex numbers. These are several relations among $X_i$, $Y_i$, $Z_i$
from the requirement of unitarity of the mixing matrix \cite{MHDM}.

The charged Higgs interactions in Eq.(\ref{chargedhiggs}) can 
induce extra contribution to FCNC processes in this model. Here, we
are interested in FCNC processes related to the $b$ quark. 
Inclusive branching ratios of 
$b \rightarrow s\,\gamma$, $ b \rightarrow s\,\ell^+\ell^-$
and $b \rightarrow s\,\nu \bar{\nu}$ are calculated through 
the following weak effective Hamiltonian at the bottom scale as described 
in \cite{GSW,Ali};
\begin{eqnarray}
  \label{H_eff}
  {\cal H} = -\frac{4G_F}{\sqrt{2}}V_{tb}V_{ts}^\ast\sum_{i=1}^{10} C_i O_i .
\end{eqnarray}
The relevant operators at the bottom scale are
\begin{eqnarray}
  \label{operator}
  O_7  &=& \frac{e}{16\pi^2}\,m_b\,{\overline s}_L \sigma_{\mu\nu} b_R
        F^{\mu\nu} ,
\\
  O_8  &=& \frac{g_s}{16\pi^2}\,m_b\,{\overline s}_L T^a \sigma_{\mu\nu} b_R
          G^{a\mu\nu} ,
\\
  O_9  &=& \frac{e^2}{16\pi^2}
  {\overline s}_L \gamma^\mu b_L \, {\overline \ell} \gamma_\mu
          \ell ,
\\
  O_{10}&=& \frac{e^2}{16\pi^2}
{\overline s}_L \gamma^\mu b_L \, {\overline \ell} \gamma_\mu
          \gamma^5 \ell ,
\\
  O_{11}&=& \frac{e^2}{16\pi^2\sin^2\theta_W}
   {\overline s}_L \gamma^\mu b_L \, \sum_{i=e,\mu,\tau}
  {\overline \nu_i} \gamma_\mu
          (1-\gamma^5) \nu_i ,
\end{eqnarray}
where $e$, $g_s$ are QED and strong coupling constants respectively,
and $\theta_W$ is the weak mixing angle.
Using renormalization group equations of QCD the Wilson coefficients
$C_i$'s at the bottom scale are related to those at the weak scale.
New physics effects enter through the Wilson coefficients at the
weak scale. From the charged Higgs 
interactions these coefficients receive the following new contributions,
\begin{eqnarray}
  \label{wilson}
  C_7^H &=& -\frac{1}{2} \sum_{i=1}^{n-1}\,x_{th_i}
          \left[ |Y_i|^2 
                \left( 
                  \frac{2}{3} F_1(x_{th_i}) + F_2(x_{th_i})
                \right)
          \right.
\\ \nonumber
          &&\left.
               + X_i Y_i^{\ast}
                \left(
                  \frac{2}{3} F_3(x_{th_i}) + F_4(x_{th_i})
                \right)
          \right],
\\
  C_8^H &=& -\frac{1}{2} \sum_{i=1}^{n-1}\,x_{th_i}
          \left[ 
                |Y_i|^2 F_1(x_{th_i}) + X_i Y_i^{\ast} F_3(x_{th_i})
          \right],
\\
  C_9^H &=& -D^H + \frac{1-4\sin^2\theta_W}{\sin^2\theta_W} C^H ,
\\
  C_{10}^H &=& -\frac{C^H}{\sin^2\theta_W},
\\
  C_{11}^H &=& -C^H.
\end{eqnarray}
Here $C^H$ and $D^H$ are given by
\begin{eqnarray}
  \label{function}
  C^H &=& \sum_{i=1}^{n-1} \frac{1}{8} |Y_i|^2 x_{th_i} x_{tW}
        \left( F_3(x_{th_i}) + F_4(x_{th_i}) \right) ,
\\
  D^H &=& \sum_{i=1}^{n-1} x_{th_i} |Y_i|^2
        \left(\frac{2}{3} F_5(x_{th_i}) - F_6(x_{th_i}) \right) ,
\end{eqnarray}
where $x_{tW} = m_t^2/m_W^2, x_{th_i} = m_t^2/m_{H_i}^2$.
The functions $F_1$-$F_6$ are defined as follows:
$F_1(x)=\frac{1}{12(x-1)^4}(x^3-6x^2+3x+2+6x\ln x)$,
$F_2(x)=\frac{1}{12(x-1)^4}(2x^3+3x^2-6x+1-6x^2\ln x)$,
$F_3(x)=\frac{1}{2(x-1)^3}(x^2-4x+3+2\ln x)$,
$F_4(x)=\frac{1}{2(x-1)^3}(x^2-1-2x\ln x)$,
$F_5(x)=\frac{1}{36(x-1)^4}(7x^3-36x^2+45x-16+(18x-12)\ln x)$,
$F_6(x)=\frac{1}{36(x-1)^4}(-11x^3+18x^2-9x+2+6x^3\ln x)$.
The Wilson coefficients at the weak scale are then given by
\begin{eqnarray}
  \label{C_i}
  C_i (m_W) = C_i^{SM} (m_W) + C_i^H (m_W) ,
\end{eqnarray}
where $C_i^{SM}$ is contribution from the SM \cite{GSW,Ali}. After taking
account of the QCD corrections at the leading logarithmic order,
$C_i(m_b)$ can be expressed as,
\begin{eqnarray}
  \label{c_mb}
  C_7(m_b) &=& C_7(m_W)\eta^{\frac{16}{23}} 
             + C_8(m_W) \frac{8}{3}(\eta^{\frac{14}{23}} 
                          - \eta^{\frac{16}{23}})
             + {\tilde C}_7 ,
\\
  C_8(m_b) &=& C_8(m_W) \eta^{\frac{14}{23}} + {\tilde C}_8 ,
\\
  C_9(m_b) &=& C_9(m_W) + {\tilde C}_9 ,
\\
  C_{10}(m_b) &=& C_{10}(m_W) ,
\\
  C_{11}(m_b) &=& C_{11}(m_W) ,
\end{eqnarray}
where $\eta = \alpha_s(m_W)/\alpha_s(m_b)$. 
${\tilde C}_7 \sim {\tilde C}_9$ are constants which depend on the QCD
coupling constant. Detailed formulas are found in \cite{Ali,GOST}.
Numerically these are given by
${\tilde C}_7=-0.17$, ${\tilde C}_8=-0.077$, ${\tilde C}_9=1.9$
for $\alpha_s(m_Z)=0.12$. 

The $b\rightarrow s\,\gamma$ branching ratio
is then given by
\begin{eqnarray}
  \label{brbsg}
  {\rm Br}(b\rightarrow s\,\gamma) = 
  \br(b \rightarrow c\ e \ \overline{\nu})
  \frac{6 \alpha_{em}}{\pi g(m_c/m_b)}
        \left|\frac{V_{ts}V_{tb}^{\ast}}{V_{cb}}\right|^2
        |C_7(m_b)|^2,
\end{eqnarray}
where the phase space factor $g(z)$ is given by 
$g(z) = 1-8z^2+8z^6-z^8-24z^4\ln z$. 
For the $b\rightarrow s\,\ell^+\ell^-$ process, the differential
branching ratio and the forward-backward asymmetry of leptons
in the lepton-center-of-mass frame are given by
\begin{eqnarray}
  \label{branch}
 && \frac{d \br(b \rightarrow s\,\ell^+\ell^-)}{d\,{\hat{s}}}
 = \br(b \rightarrow c\,e{\overline \nu}) \frac{\alpha_{em}^2}{4\pi}
        \left|\frac{V_{ts}V_{tb}^{\ast}}{V_{cb}}\right|^2
        \frac{1}{g(m_c/m_b)}(1-\hat{s})^2
\\ \nonumber
 &&\times \left[ (|C_9 + Y(\hat{s})|^2 + |C_{10}|^2)(1+2\hat{s})
          +\frac{4}{\hat{s}}|C_7|^2(2+\hat{s})^2
          +12 \mbox{Re}C_7^\ast(C_9+Y(\hat{s}))
   \right] ,
\\
 &&A_{FB}(\hat{s})
\\ \nonumber
 &&= \frac{-3 \mbox{Re} C_{10}^\ast[ ( C_9
                +Y(\hat{s}))\hat{s}+2 C_7]}
          {(|C_9 + Y(\hat{s})|^2 + |C_{10}|^2)(1+2\hat{s})
          +\frac{4}{\hat{s}}|C_7|^2(2+\hat{s})^2
          +12 \mbox{Re}C_7^\ast(C_9+Y(\hat{s}))} ,
\end{eqnarray}
where $\hat{s}=s/m_b^2=(p_++p_-)^2/m_b^2$ and $p_+(p_-)$ is the four
momentum of $\ell^+(\ell^-)$.
$Y(\hat{s})$ represents the contribution from the charm quark loop at
the $m_b$ scale. See Ref.\cite{GOST} for details.
Here we neglect the $J/\psi$ and $\psi'$ resonance contributions.
The branching ratio for the $b \rightarrow s\,\nu\overline{\nu}$ is
given by
\begin{eqnarray}
  \label{bsnn}
\sum_{i=e,\mu,\tau}\br(b\rightarrow s\, \nu_i \overline{\nu}_i)
=  3\ \br(b \rightarrow c\ e \ \overline{\nu})
        \frac{\alpha^2}{4 \pi^2\sin\theta_W^4}
        \left|\frac{V_{tb}V_{ts}^\ast}{V_{cb}}\right|^2
         \frac{1}{g(m_c/m_b)}|C_{11}|^2.
\end{eqnarray}

Using the above formulas, it is now straight-forward to evaluate
the branching ratios and the asymmetry numerically
 in the multi-Higgs doublet model.
In the followings, we assume that 
only one of the physical charged Higgs boson is light and neglect the
effects of other physical charged Higgs bosons. Then we keep only one
term in the summation in the expressions for $C_i^H$'s. Dropping the
index $i$ for the lightest charged Higgs bosons, the relevant parameters
are $|Y|^2$, $XY^\ast$ and the mass of the charged Higgs boson.
It should be noted that
the coefficients $C_7$ and $C_8$ depend on both $|Y|^2$ and $XY^\ast$ 
whereas $C_9$ and $C_{10}$ only contain $|Y|^2$. 

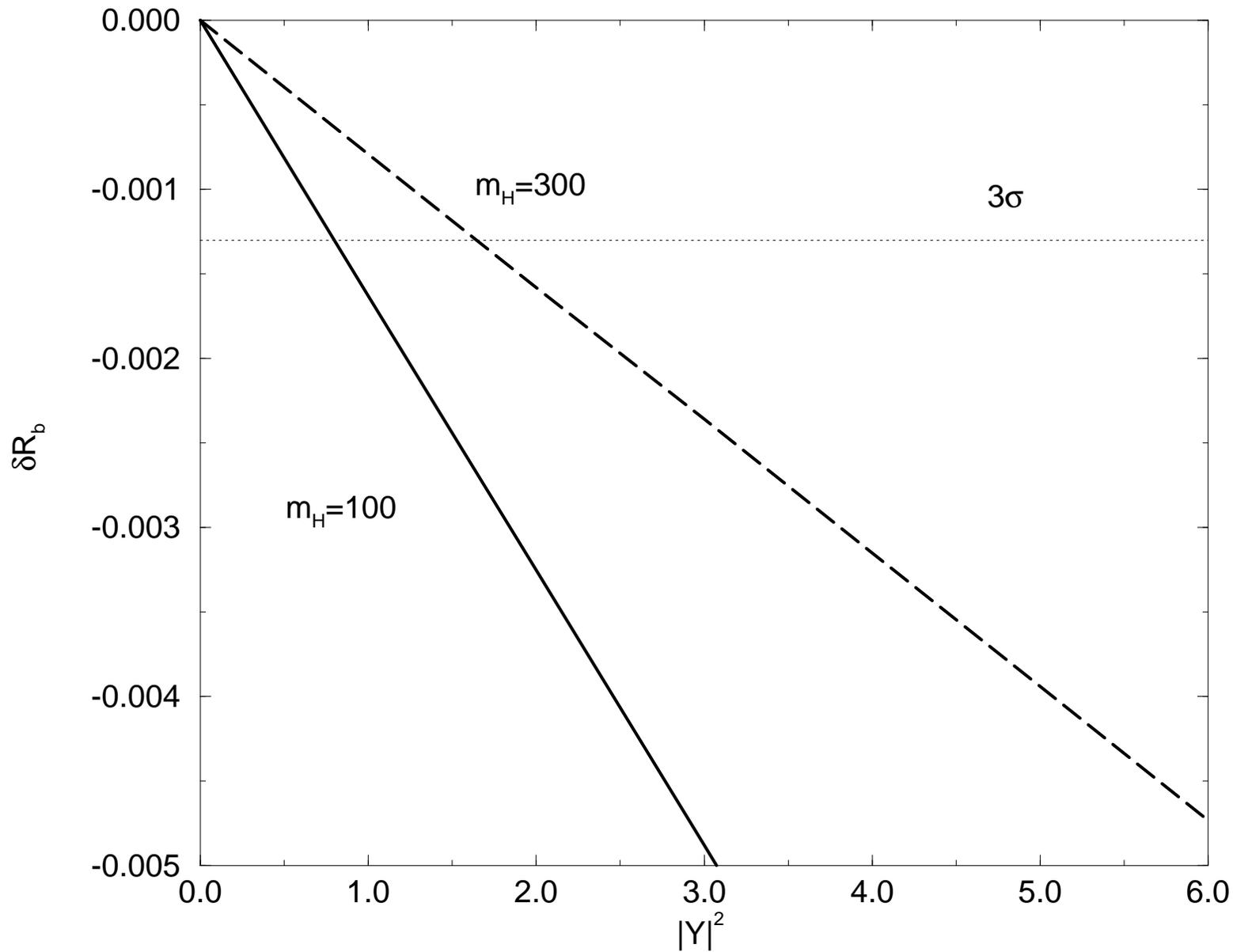
\begin{figure}[htbp]
  \begin{center}
    \leavevmode
    
    \caption{The charged Higgs contribution to $\delta R_b$
      for $m_H=100$ GeV(solid line) and $m_H=300$ GeV(dashed line).
      The dotted line represents the 3 $\sigma $ lower bound
      from the world average value of $R_b$.}
    \label{fig:Rb}
  \end{center}
\end{figure}

Since the $b \rightarrow s\,\gamma$ branching ratio is already observed
and is consistent with the SM prediction,
we use $\br(b\rightarrow s\,\gamma)$ to solve $\mbox{Im}XY^\ast$ in
terms of $|Y|^2$ and $\mbox{Re}XY^\ast$ for each value of the charged
Higgs mass.
Then other observable quantities for $b \rightarrow s\,\ell^+\ell^-$ and
$b \rightarrow s\,\nu\overline{\nu}$ processes
can be calculated as functions of two parameters.
We should also take into account constraints from other processes.
Besides the 
$b\rightarrow s\,\gamma$ process,
the $B^0$-${\overline B}^0$ mixing and the $Z \rightarrow b\overline{b}$
process give the most strong constraints on
the possible value of $|Y|^2$ \cite{MHDM}. The contribution to the 
$B^0$-${\overline B}^0$ mixing from the charged Higgs boson is expressed
as 
\begin{eqnarray}
  \label{BBbar}
  M_{12}^H &=& \frac{G_F^4}{64\pi^2}m_W^2 \eta_B (V_{td}V_{tb}^\ast)^2
             \frac{4}{3} B_B f_B^2 m_B x_t\left[ y_t I_1(y_t) |Y|^4
\right.
\\ \nonumber
&&\left.
             + x_t (2I_2(x_t,x_H)-8I_3(x_t,x_H) |Y|^2)
\right],
\end{eqnarray}
where $y_t = m_t^2/m_H^2$, $x_H = m_H^2/m_W^2$,
and the functions $I_1$-$I_3$ are defined as follows.
\begin{eqnarray}
  \label{func2}
 I_1(x) &=&\frac{1+x}{(1-x)^2}+ \frac{2 x \ln x}{(1-x)^3},
\\ 
I_2(x,y)&=&\frac{x}{(x-y)(x-1)}+\frac{y^2\ln(y)}
{(y-1)(x-y)^2}+\frac{x(-x-xy+2y)\ln(x)}{(1-x)^2(x-y)^2},
\\
I_3(x,y)&=&\frac{1}{(x-y)(x-1)}+\frac{y\ln{y}}{(y-1)(x-y)^2} 
+\frac{(-x^2+y)\ln(x)}{(1-x)^2(x-y)^2}.
\end{eqnarray}
Here we have retained only the relevant terms which are proportional
to $|Y|^4$ and $|Y|^2$. Since this quantity depends on the CKM
matrix element $V_{td}$ which has not been known well, the constraint
from $B^0$-${\overline B}^0$ mixing is not very strong. Using the
constraint on $V_{td}$ from the charmless $b$ decay ($0.005<|V_{td}|<0.012$)
and taking account of uncertainties from $f_B$ and $B_B$ 
($f_B\sqrt{B_B}=200\pm40$MeV)
we can deduce $|Y| \lsim 1.5(2.1)$ 
for $m_H=100(300)$ GeV
\footnote{
The CP violating parameter in the $K^0$-${\overline K}^0$ mixing,
$\epsilon_K$, also receives a similar contribution from the charged
Higgs loop as the $B^0$-${\overline B}^0$ mixing. In this case, however,
the relevant CKM matrix element is different and the constraint on
$|Y|$ from $\epsilon_K$ only is not strong. If we combine the
$B^0$-${\overline B}^0$ mixing and the $\epsilon_K$ constraints we can
exclude a slightly larger parameter space but the allowed region of $|Y|$ is 
numerically almost the same as above.}.
Here and in the followings we fix the top quark mass as 175 GeV.
For the $Z\rightarrow b\overline{b}$ process we calculated the charged Higgs
contribution to 
$R_b=\Gamma(Z\rightarrow b\overline{b})/
\Gamma(Z\rightarrow {\rm hadrons})$ following Ref.\cite{Rb}.
We define the deviation from the SM contribution ($R_b^{SM}= 0.2158$)
by $\delta R_b = R_b - R_b^{SM}$. The charged Higgs contribution to
$\delta R_b$ is shown in Fig.\ref{fig:Rb} as a function of $|Y|^2$.
In this figure we also show the 3 $\sigma$ lower bound of $\delta R_b$
from the world average value $R_b= 0.2178 \pm 0.0011$ \cite{world}.
From Fig.\ref{fig:Rb} we can derive $|Y| \lsim 0.9(1.3)$
for $m_H=100(300)$ GeV.
Note that if we use the value $R_b = 0.2159 \pm 0.0009({\mbox stat}) \pm 
0.0011({\mbox syst})$ 
reported by ALEPH \cite{ALEPH}
the lower bound of $\delta R_b$ shifts to $-0.0041$, which corresponds
to the upper bound $|Y| \lsim 1.6(2.3)$
for $m_H=100(300)$ GeV.
Since experimental situation is not conclusive,
to be conservative,
we present the result in a rather wider range of $|Y|$.

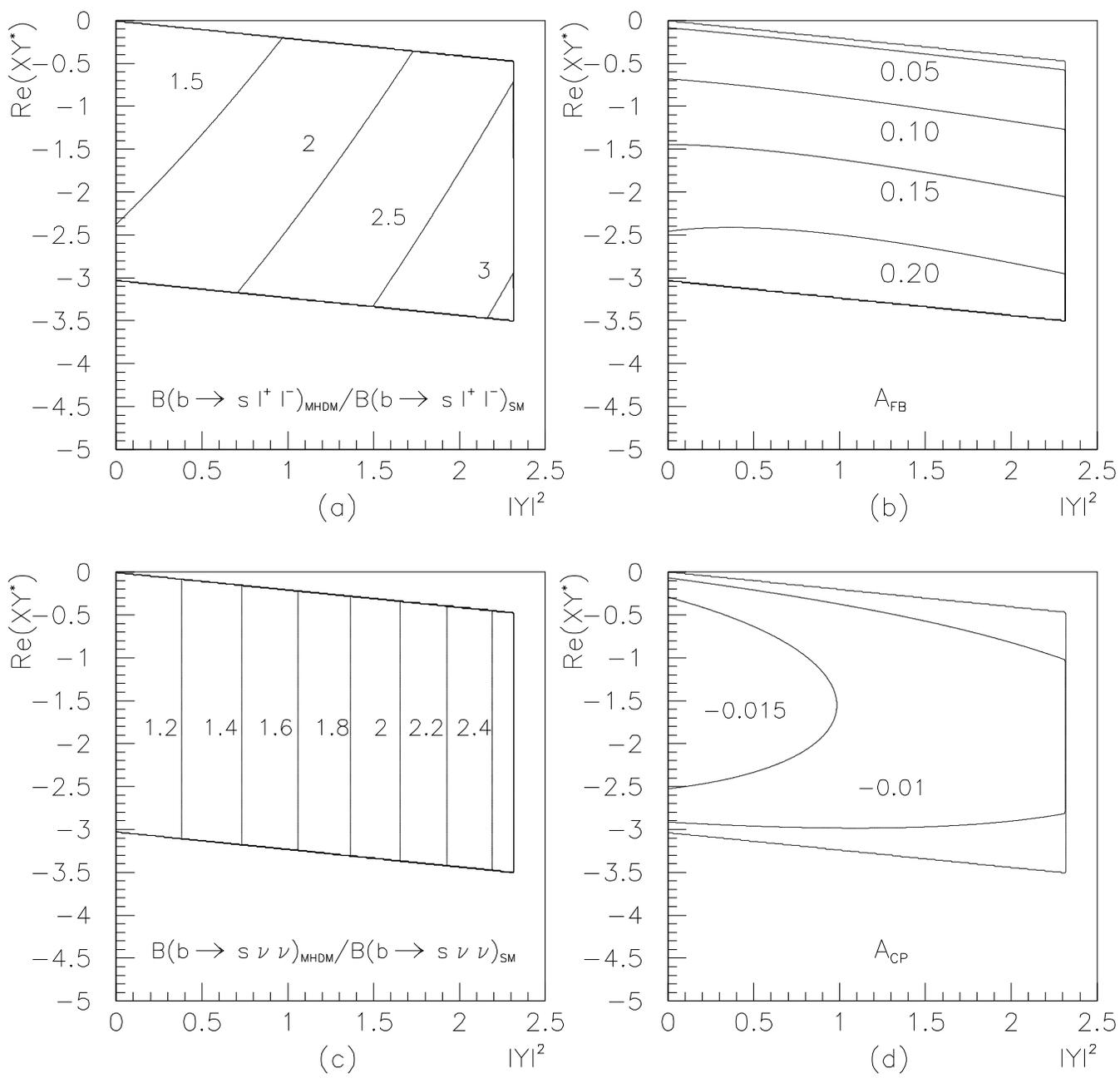
\begin{figure}[htbp]
  \begin{center}
    \leavevmode
    \caption{(a).The ratio of the $\br( b \rightarrow \,s \ell^+\ell^-)$ in
      the range $4m_\mu^2 \leq s \leq (m_{J/\psi}-\delta)^2$ for
      the multi-Higgs doublet model(MHDM) and the standard model(SM)
      in the case of $m_H=100$ GeV.
      (b).The forward-backward asymmetry$(A_{FB})$ in
      the range $4m_\mu^2 \leq s \leq (m_{J/\psi}-\delta)^2$ for
      the MHDM and the SM in the case of $m_H=100$ GeV.
      (c).The ratio of the $\br( b \rightarrow \,s \nu\overline{\nu})$
      for the MHDM and the SM in the case of $m_H=100$ GeV.
      (d).The charge asymmetry$(A_{CP})$ in the range 
      $(m_{\psi'} + \delta)^2 \leq s \leq m_b^2$ for $m_H=100$ GeV.
      We take $\delta=100$ MeV in (a), (b) and (d).
 }
    \label{fig:1}
  \end{center}
\end{figure}

Fig.\ref{fig:1}(a) shows that the ratio of 
$\br(b \rightarrow s\,\mu^+\mu^-)$ branching ratio
for the multi-Higgs doublet model normalized by the SM 
branching ratio in the space of
$|Y|^2$ and $\mbox{Re}(XY^\ast)$ for $m_H=100$ GeV.
For numerical calculations we use $m_b=4.7$ GeV and $m_c=1.5$ GeV.
To avoid the large effect of the $J/\psi$ resonance we
integrate the differential branching ratio for the kinematical range 
$4m_\mu^2 < s < (m_{J/\psi}-\delta)^2$ where $m_\mu$ is muon mass,
$m_{J/\psi}$ is $J/\psi$ mass and $\delta=100$ MeV.
The interference effect between the $J/\psi$ resonance and the short
distance contribution in this kinematical region is still
sizable $(\lsim 20 \%)$ for the SM case \cite{GOST}. However as we can
see in Fig.\ref{fig:1} the interference effect is expected to be
smaller than the charged Higgs contribution to the short distance part.
In order to solve $\mbox{Im}(XY^\ast)$ in terms of $|Y|^2$ and
$\mbox{Re}(XY^\ast)$,
we assume $\br(b\rightarrow s\,\gamma)= 2.8 \times 10^{-4}$.
Since there is a sizable experimental error on this quantity we have just
used the SM prediction for the illustration.
We expect that the experimental error as well as the theoretical ambiguity
will be reduced in future when actual analysis on 
$b \rightarrow s\,\ell^+\ell^-$ is
done. 
Fig.\ref{fig:1}(b) shows the asymmetry in the same kinematical range.
From these two figures we can see that the branching ratio can be
a few times lager than that of the SM, which is 
$3.8 \times 10^{-6}$ in this kinematical region, and the asymmetry can
be as large as 20 \% compared to about 5 \% in the SM.
It is also interesting to see that the branching ratio is sensitive to
the value of $|Y|^2$, on the other hand the asymmetry gives an
independent constraint on the parameter space. Therefore
the values of $|Y|^2$ and $\mbox{Re}(XY^\ast)$ are determined
if the branching ratio and the asymmetry are measured with
reasonable accuracy.
Fig.\ref{fig:1}(c) shows the $b\rightarrow s\,\nu\overline{\nu}$
branching ratio normalized by the SM prediction 
($\br(b\rightarrow s\,\nu\overline{\nu})_{SM}=4.3\times 10^{-5}$)
in the same parameter space. 
In this case the branching ratio only depends on the parameter $|Y|^2$
and is enhanced by a similar factor as the branching ratio of 
$b \rightarrow s\,\ell^+\ell^-$.

If $\mbox{Im}(XY^\ast) \neq 0$, there is possibility to observe the CP
violating charge asymmetry, \ie the difference of 
$b \rightarrow s\,\ell^+\ell^-$ and 
${\overline b} \rightarrow \overline{s}\,\ell^+\ell^-$.
Since this quantity is induced with help of the phase in 
$Y(\hat{s})$ in Eq.(\ref{branch}), the asymmetry appears only
above the $c{\overline c}$ threshold of the lepton invariant mass.
The charge asymmetry is defined as 
\begin{eqnarray}
  \label{CP}
  A_{CP} =
        \frac{\int_{(m_{\psi'}+\delta)^2}^{m_b^2} ds
              \left(\frac{d\,\br(b \rightarrow s\,\ell^+\ell^-)}{ds}
                    -\frac{d\,\br({\overline b} \rightarrow {\overline s}
                      \,\ell^+\ell^-)}{ds}
              \right)}
             {\int_{(m_{\psi'}+\delta)^2}^{m_b^2} ds
              \left(\frac{d\,\br(b \rightarrow s\,\ell^+\ell^-)}{ds}
                    +\frac{d\,\br({\overline b} \rightarrow {\overline s}
                      \,\ell^+\ell^-)}{ds}
              \right)} ,
\end{eqnarray}
where $m_{\psi'}$ is the $\psi'$ mass and we take $\delta = 100 $ MeV.
In Fig.\ref{fig:1}(d) this quantity is shown 
for $m_H=100$ GeV.
Since this asymmetry is at most a few percent, we need large statistics
to measure it. Note that there is a twofold ambiguity in sign of
$\mbox{Im}(XY^\ast)$. Fig.\ref{fig:1}(a)-(c) do not depend on the sign
of $\mbox{Im}(XY^\ast)$ while $A_{CP}$ (Fig.\ref{fig:1}(d)) change its
sign according to the sign of $\mbox{Im}(XY^\ast)$. Fig.\ref{fig:1}(d)
shows the case of $\mbox{Im}(XY^\ast) > 0$.

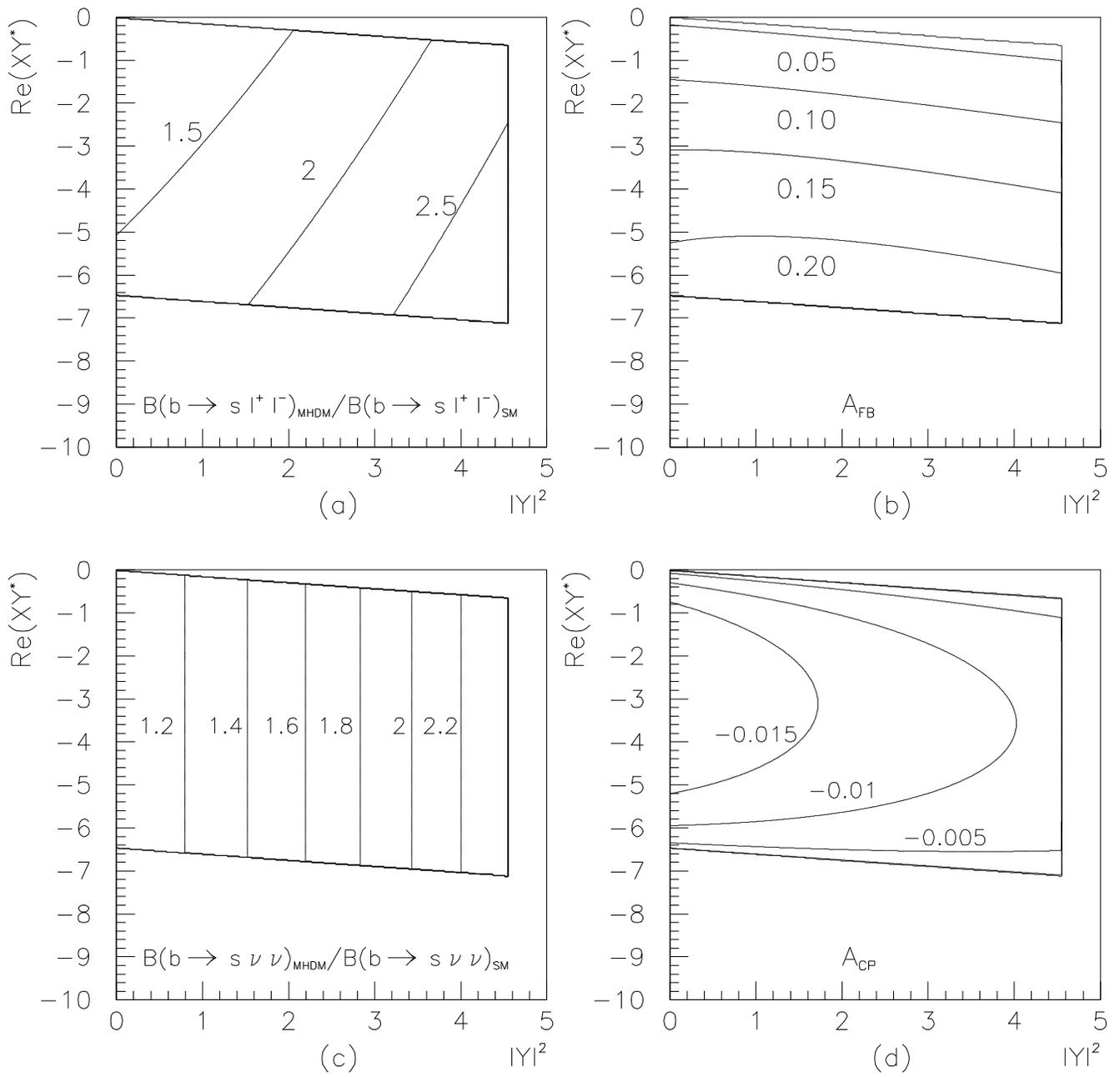
\begin{figure}[htbp]
  \begin{center}
    \leavevmode
    \caption{The same as \reffig{fig:1} for $m_H=300$ GeV.}
    \label{fig:2}
  \end{center}
\end{figure}

In Fig.\ref{fig:2}(a)-(d) we show the same quantities for
$m_H=300$ GeV. As in the case of $m_H=100$ GeV we can see similar
enhancements on the branching ratios 
and the forward-backward asymmetry in the allowed parameter space.

In summary we have investigated possible constraints on the charged
Higgs coupling parameters from the $b \rightarrow s\,\gamma$,
$b \rightarrow s\,\ell^+\ell^-$  and $b \rightarrow s\,\nu\overline{\nu}$
processes. 
Within the present experimental constraints including 
$b \rightarrow s\,\gamma$ process, we have shown that the branching
ratio of the $b \rightarrow s\,\ell^+\ell^-$ 
and $b \rightarrow s\,\nu\overline{\nu}$ processes can be enhanced a few
times compared with the SM. The lepton forward-backward asymmetry in the 
$b \rightarrow s\,\ell^+\ell^-$ process can also be enhanced by 
a factor of several.
Since these quantities depend on the model parameters in different ways,
we can obtain useful information on the model once these processes are
observed experimentally.

This work is supported by the Grant-in-aid for Scientific Research from
the Ministry of Education, Science and Culture of Japan.

\newpage

\end{document}